\documentclass{aa}  

\usepackage{graphicx}
\usepackage{txfonts}
 \usepackage[colorlinks=true,     linkcolor=blue, citecolor=blue, filecolor=blue, urlcolor=blue]{hyperref}

\newcommand{\vect}{\overrightarrow}

\bibpunct{(}{)}{;}{a}{}{,} 

\newcommand{\msgguy}[1]{#1} 
\newcommand{\msgref}[1]{#1} 
\newcommand{\redguy}[1]{#1}

\usepackage{relsize}
\newcommand{\smallletter}[1]{\mathsmaller{\mathsmaller{#1}}}
\newcommand{\stimes}{\smallletter{\times}}

\begin{document}

\title{Theory of optical long-baseline interferometry on polarized sources}
   \author{G. Perrin
          \inst{1}\thanks{\redguy{formerly LESIA, Observatoire de Paris, Universit\'e PSL, CNRS, Sorbonne Universit\'e, Universit\'e Paris Cit\'e, 5 place Jules Janssen, 92195 Meudon, France}}
          }
    \institute{\redguy{LIRA}, Observatoire de Paris, Universit\'e PSL, CNRS, Sorbonne Universit\'e, Universit\'e Paris Cit\'e, 5 place Jules Janssen, 92195 Meudon, France\\
              \email{guy.perrin@obspm.fr}
             }

   \date{Received December 24, 2025; accepted February 19, 2026}

 
  \abstract 
{The effects of the polarization characteristics of beam trains in optical long-baseline interferometers are well known and have led to difficulties in measuring the spatial coherence of astronomical sources in the past. This has been overcome by designing symmetrical optical trains. With the advent of interferometers using large telescopes, observations of faint sources with high degrees of polarization have become \msgguy{even more possible}. As in the radio domain, where radiation processes usually lead to high polarization rates, a description of coherence for polarized or unpolarized sources observed with non-polarization neutral interferometers is necessary. A theory of optical long-baseline interferometry fully taking into account the polarization characteristics of beam trains and those of the sources is presented in this paper, building on concepts developed for radio aperture synthesis. \msgguy{The concept of generalized Mueller matrix is introduced for the case of multi-aperture interferometers leading to a} simple matrix relationship between the \msgguy{observed Stokes visibilities, as they are disturbed by the instrument polarization characteristics,} and \msgguy{the} object \msgguy{Stokes} visibilities. \msgguy{This relationship is} applied \msgguy{to the case of} single-mode interferometers. \msgref{The formalism also shows that classical complex visibilities (squared moduli, phases and closure phases) need to be debiased from polarization crosstalk, even when the source is not polarized as in this case ghost polarized visibilities are created.} }

   \keywords{instrumentation: high angular resolution -- instrumentation: interferometers -- methods: analytical -- techniques: polarimetric}

   \maketitle

%
%

\section{Introduction}
Long-baseline interferometers measure spatial coherence between distant sampling points. Although the basis of theory for long-baseline interferometry is the same for the radio and optical domains, there are very important differences in their practical implementation and in the processing of data to get images. Because radio waves can be directly measured at the focus of antennas, there is no need to propagate beams down to a common recombination point as is the case in the optical domain, making optical interferometers more complex systems than radio interferometers. 
A consequence of this important difference is that the polarizing properties of optical interferometers can be very complex, with a few tens of optical components that reflect or transmit the beams. This had huge consequences in the past for the ability to detect fringes, as was experienced by Labeyrie before his first historical detection of long-baseline fringes at optical wavelengths \citep{Labeyrie1975}. This also had consequences for how optical interferometers are designed to avoid the effects of differential birefringence between the beams that can kill fringe contrast, of differential rotation between polarization planes, or of differences in coatings, for example, which can have the same effect \citep{Traub1988,Perraut1996}: interferometers need to be symmetric, which still increases the number of optical reflections. These effects also have consequences for the process to measure complex visibilities on polarized sources especially as it is now possible to observe \msgguy{even more} sources with high degrees of polarization thanks to the access to much fainter sources where radiation can be produced by processes such as synchrotron radiation (see for example the results of GRAVITY on flares at the Galactic Center in \citet{GRAVITY_flares_2018}). Some authors have addressed these issues in the past, including \citet{Elias2001,Elias2004} and \citet{Buscher2009}, with Jones and Mueller matrices to propagate polarization states and Stokes parameters across the interferometer, or more recently \citet{Widmann2023}, who studied the polarization characteristics of the  Very Large Telescope Interferometer (VLTI) and their effects. \msgref{\citet{Shuai2025} recently carried out a similar work as in the present paper and applied it to the MIRC-X and MYSTIC instruments of the CHARA interferometer}. These works use different notations or definitions, even sometimes for similar quantities. The goal of this paper is to establish a new comprehensive formalism for optical long-baseline interferometry generalized to polarizing interferometers and polarized sources, partly based on what has been done for decades in the radio domain (see for example the  monograph on radio interferometry and aperture synthesis by \citet{Thompson2017} for a complete description). \\
The paper is organized as follows: the basic formalism of coherence in an interferometer for polarized waves is given in Section~\ref{sec:formalism}, the propagation of coherence taking into account the polarizing properties of an interferometer is addressed in Section~\ref{sec:propagation}, and the consequences for the calibration of complex visibilities for single-mode interferometers is described in Section~\ref{sec:calibration}. Examples of instrumental polarization effects are given in Section~\ref{sec:examples} as are examples of coherence for some specific sources. Some conclusions are given in the last section.

%
%

\section{Formalism of the polarized waves and coherence in an interferometer}
\label{sec:formalism}
In this paper, monochromatic polarized waves propagate in vacuum and are described as complex vectors called Jones vectors of the form
\begin{equation}
\label{eq:jones_vector}
 \vect{E}(\vect{r},\lambda,t) = e^{-i(\vect{k}.\vect{r} -\omega t)}\!\left[\begin{array}{ll}
                      \!\!\!\,E_x \\
                      \!\!\!\,E_y
                     \end{array} \!\!\!\right] 
,\end{equation}
where $x$ and $y$ are the two orthogonal axes along which polarizations are split or analyzed in the beam-combining instrument. The wave vector, $\vect{k}$, is pointing in the direction of propagation and its norm is equal to $2\pi/\lambda$. The pulsation, $\omega$, relates to the wavelength, $\lambda$, through $\omega = 2\pi c /\lambda$. Monochromatic waves are not strictly speaking physical and should be integrated across a finite bandpass to define quasi-monochromatic waves in the sense of \citet{Goodman} with the consequence of systematically having an additional integral in all the equations to follow. The choice is made to consider monochromatic waves only and ignore temporal coherence effects. In case these effects are important to consider for a specific application, all waves and correlations can be integrated across a given bandwidth {a posteriori}. Dependency on wavelength is therefore implicit in what follows.
Since only spatial coherence is addressed here, correlations between waves are systematically measured at the same times and therefore with no differential time dependency. As a consequence, the  $e^{-i\omega t}$ term is also dropped from now on. \\
The waves are emitted by astronomical sources with some spatial extension. The above Jones vector is therefore the integral of a spatial distribution of Jones vectors dependent on astronomical coordinates, $\alpha$ and $\delta$, over the spatial extension of the source:
\begin{equation}
\label{Eq:coherence_Kronecker}
\vect{E}(\vect{r}) =\iint_{source}\vect{\mathcal{E}}(\alpha,\delta)d\alpha d\delta
,\end{equation}
where the cursive style is used for the spatial wave distribution. In what follows, the limitation to the source spatial extension is always implicit. In practice it is also limited by the field of view of the instrument or by the fundamental mode in the case of a single-mode interferometer, as is also the case in radio. The waves are now expanded in unpolarized and polarized components:
\begin{equation}
 \vect{E}(\vect{r}) = e^{-i\vect{k}.\vect{r}}\!\left[\begin{array}{ll}
                      \!\!\!\,E_{np,x}+E_{p,x} \\
                      \!\!\!\,E_{np,y}+E_{p,y}
                     \end{array} \!\!\!\right] 
.\end{equation}
This is done for all beams in an interferometer identified by a superscript, $n$, in $\vect{E^n}(\vect{r})$ for beam $n$. Long baseline interferometry is about measuring the spatial coherence of the source at the points where the waves are collected by telescopes. This coherence is mixed with polarization coherence and both can be characterized by the coherence matrix of \citet{born}:
 \begin{equation}
 \label{Eq:coherence_matrix}
                C=\left[\begin{array}{ll}
                      \!\!\!<E_xE_x^{\star}> & <E_xE_y^{\star}> \!\!\!  \\
                      \!\!\!<E_yE_x^{\star}> & <E_yE_y^{\star}>\!\!\!
                     \end{array} \right] 
,\end{equation}
where the superscript $\star$ is the complex conjugation and the terms of the matrix are correlations between the wave components. Note that this matrix can also be written as the average product of the wave vector by its conjugate transpose noted with the dagger:
\begin{equation}
\label{Eq:coherence_product}
C = <\vect{E}.\vect{E}^\dag\!\!>
.\end{equation}
In the optical, the terms of the matrix can be measured by splitting polarizations with a Wollaston prism ($<\!\!E_xE_x^{\star}\!\!>$ and $<\!\!E_yE_y^{\star}\!\!>$), by rotating the polarizations by $45^o$ with a half-wave plate before splitting (\msgguy{$<\!\! E_xE_y^{\star}+E_yE_x^{\star}\!\!>$}), and by applying  a quarter wave-plate rotated by $45^o$ before splitting (\msgguy{$<\!\! E_xE_y^{\star}-E_yE_x^{\star}\!\!>$}). The total flux from the source is obtained by taking the trace of the coherence matrix $I=\mathrm{tr}(C)=
<\!\!|E_x|^2\!\!>+<\!\!|E_y|^2\!\!>$. The polarized intensity is $PI=<\!\!|E_{p,x}|^2\!\!>+<\!\!|E_{p,y}|^2\!\!>$, with $P$, the degree of polarization. 

Since the polarized and unpolarized waves are not coherent between each other and since the $x$ and $y$ components of the unpolarized waves are incoherent, the matrix can also be written:
 \begin{equation}
 \begin{split}
                C=\left[\begin{array}{ll}
                      \!\!\!<E_{np,x}E_{np,x}^{\star}> & 0 \!\!\!  \\
                      \!\!\!0& <E_{np,y}E_{np,y}^{\star}>\!\!\!
                     \end{array} \right]  & \\
           &  \!\!\!\!\!\!\!\!\!\!\!\!\!\!\!\!\!\!\!\!\!\!\!\!\!\!\!\!\!\!\!\!\!\!\!\!\!\!\!\!\!\!\!\!\!\!\!\!+
                     \,\,\,\,\, \,\,\,\,\, \,\,\,\,\, \,\,\,\,\,\left[\begin{array}{ll}
                      \!\!\!<E_{p,x}E_{p,x}^{\star}> & <E_{p,x}E_{p,y}^{\star}> \!\!\!  \\
                      \!\!\!<E_{p,y}E_{p,x}^{\star}> & <E_{p,y}E_{p,y}^{\star}>\!\!\!
                     \end{array} \right] 
\end{split}
.\end{equation}
For the unpolarized part of the wave, the first matrix reads as
\begin{equation}
                C_{np}=(1-P)\frac{I}{2} \stimes I_2
,\end{equation}
with $I_2$ the 2x2 identity matrix. The second matrix can be written with the Stokes parameters $Q$, $U$, and $V$:
 \begin{equation}
                C_{p}= \frac {1}{2}\left[\begin{array}{ll}
                      \!\!\!PI+Q& \msgref{U+iV} \!\!\!  \\
                      \!\!\!\msgref{U-iV} &PI-Q\!\!\!
                     \end{array} \right] 
.\end{equation}
\msgref{The definition used for the Stokes parameters is the the one of \cite{Hamaker_Bregman1996}. It is aligned with the IAU convention of 1974 when $x$ is the declination axis increasing toward the north and $y$ the right ascension axis increasing toward the east, and takes into account the definition of the harmonic exponential in Eq.(\ref{eq:jones_vector}) for the sign and expression of the $V$ Stokes parameter. \citet{Shuai2025} have chosen different conventions; hence, the differences in signs in the coherence matrix.} Note that the trace of this part of the coherence matrix is equal to the polarized intensity: $\mathrm{tr}(C_p)=PI$. The sum of the two matrices equates to half the Stokes brightness matrix of \citet{Hamaker2000}:
%
 \begin{equation}
                C= \frac{1}{2}\left[\begin{array}{ll}
                      \!\!\!I+Q& \msgref{U+iV} \!\!\!  \\
                      \!\!\!\msgref{U-iV} &I-Q\!\!\!
                     \end{array} \right] 
.\end{equation}
Stokes parameter spatial distributions can also be defined assuming the emitter is spatially incoherent, the products of double integrals in Eq.(\ref{Eq:coherence_matrix}) being replaced by a single double integral of the products of wave components as was done in \citet{Goodman}:
\begin{equation}
 \left\{\begin{array}{@{}l@{}}
\,\, I = \iint \mathcal{I}(\alpha,\delta)d\alpha d\delta \,\,\,= \!\! \iint \!\!<\!\mathcal{E}_x\mathcal{E}_x^{\star}(\alpha,\delta)+\mathcal{E}_y\mathcal{E}_y^{\star}(\alpha,\delta)\!\!>d\alpha d\delta\\ 
Q = \iint \mathcal{Q}(\alpha,\delta)d\alpha d\delta \,\,= \!\! \iint \!\!<\!\mathcal{E}_{p,x}\mathcal{E}_{p,x}^\star(\alpha,\delta)-\mathcal{E}_{p,y}\mathcal{E}_{p,y}^\star(\alpha,\delta)\!\!>d \alpha d\delta \\ 
\msgguy{U = \iint \mathcal{U}(\alpha,\delta)d\alpha d\delta = \!\! \iint \!\!<\!\mathcal{E}_{p,x}\mathcal{E}_{p,y}^\star(\alpha,\delta)+\mathcal{E}_{p,y}\mathcal{E}_{p,x}^\star(\alpha,\delta)\!\!>d\alpha d\delta}\\
\msgguy{V = \iint \mathcal{V}(\alpha,\delta)d\alpha d\delta = \!\! \iint \!\!-i\!\!<\!\mathcal{E}_{p,x}\mathcal{E}_{p,y}^\star(\alpha,\delta)-\mathcal{E}_{p,y}\mathcal{E}_{p,x}^\star(\alpha,\delta)\!\!>d\alpha d\delta}
\end{array}\right.
.\end{equation}
\msgref{These relations} can be extended to the case of an interferometer with correlations in Eq.(\ref{Eq:coherence_matrix}) replaced by cross-correlations between beams $n$ and $m$ \msgref{as was done by \cite{Hamaker_Bregman1996}}:
 \begin{equation}
 \label{Eq:coherence_matrix_nm}
                C^{nm}=\left[\begin{array}{ll}
                      \!\!\!<E_x^{n}E_x^{m\star}> & <E_x^nE_y^{m\star}> \!\!\!  \\
                      \!\!\!<E_y^{n}E_x^{m\star}> & <E_y^nE_y^{m\star}>\!\!\!
                     \end{array} \right] 
,\end{equation}
yielding
 \begin{equation}
 \label{Eq:polarized_coherence}
                C^{nm}= \frac{I}{2}V^{nm}_II_2+\frac{I}{2}\left[\begin{array}{ll}
                      \!\!\!\,\,\,\,\,\,\,\,\,\,V_{\mathcal{Q}}^{nm}& \msgref{V_{\mathcal{U}}^{nm}+iV_{\mathcal{V}}^{nm}} \!\!\!  \\
                      \!\!\! \msgref{V_{\mathcal{U}}^{nm}-iV_{\mathcal{V}}^{nm}} &\,\,\,\,\,\,\,-V_{\mathcal{Q}}^{nm}\!\!\!
                     \end{array} \right] 
,\end{equation}
where $V_{\mathcal{I}}^{nm}$ is the classical visibility defined in fundamental monographs dealing with coherence \citep{Goodman,born} and \msgguy{given} by the Zernike-van Cittert theorem. In this paper, coherence is associated with correlations of waves and is a coherent flux, while visibilities are systematically associated with normalized correlations, and therefore have no dimensions and are unitless. This is applied in the same spirit to the generalization of the visibility for the Stokes parameters $V_{\mathcal{Q}}$, $V_{\mathcal{U}}$, and $V_{\mathcal{V}}$. Note that, in the radio domain, the normalization of these quantities is different in other papers or between papers, in which the coherence matrix can be called a visibility matrix; for example, with a factor of 2 in addition (this is the case in \citet{Smirnov2011} and \citet{Hamaker2000}, for example). Also, the Stokes visibilities were first introduced by \citet{Morris1964}, in which for example the complex visibility function for the brightness distribution of $\mathcal{Q}$ was normalized by the integral of $\mathcal{Q}$, which is to say by $Q$; similar expressions were given in \citet{Conway1969} but were not normalized. There is therefore nothing new introduced at this stage in this paper except a different and consistent way to define the physical quantities used in this work. There are physical reasons for that. Because of the effect of atmospheric turbulence in the optical, the flux of photons injected in single-mode fibers is fluctuating with time and is also affected by static aberrations. The coherent flux is therefore fluctuating, while the normalized coherent flux is stable in single-mode interferometry. This is less the case when going to longer wavelengths, starting in the mid-infrared where coherence fluctuations are less severe and where it becomes possible to work with coherent fluxes as in radio. In what follows, coherent fluxes (correlations) are therefore systematically normalized with respect to total intensity to get visibilities, which is a difference compared to how the data of radio interferometers are calibrated. Hence the definitions used in this paper:
\begin{equation}
 \left\{\begin{array}{@{}l@{}}
V_{\mathcal{I}}(u,v) \,= \iint \,\,\mathcal{I}(\alpha,\delta)e^{-2i\pi(\alpha u + \delta v)}d\alpha d\delta / I\\ 
V_{\mathcal{Q}}(u,v) \,= \iint \,\,\mathcal{Q}(\alpha,\delta)e^{-2i\pi(\alpha u + \delta v)}d\alpha d\delta / I\\ 
V_{\mathcal{U}}(u,v) = \iint \mathcal{U}(\alpha,\delta)e^{-2i\pi(\alpha u + \delta v)}d\alpha d\delta / I\\
V_{\mathcal{V}}(u,v) = \iint \mathcal{V}(\alpha,\delta)e^{-2i\pi(\alpha u + \delta v)}d\alpha d\delta / I
\end{array}\right.
.\end{equation}
$(u,v)$ are the coordinates of the spatial frequencies given by the projection of $\vect{r_n}-\vect{r_m}$ on the plane of the sky in the direction of the object according to the Zernike-van Cittert theorem (a quadratic phasor needs to be introduced in the radio domain in these expressions as the mode of the antennas gives access to a much wider field of view). As for the traditional Stokes parameters, the Stokes visibilities being second-order moments are only sensitive to the differential phases between the waves. 

The coherence matrix above mixes both spatial and polarization coherences. It can be obtained without assuming that the source is point-like. It is clear from Eq.(\ref{Eq:polarized_coherence}) that the classical visibility, $V^{nm}_{\mathcal{I}}$, is biased by the polarization characteristics of the source if these are not taken into account to properly measure it. The 
Stokes visibilities of the object can be fully determined by inversion of Eq.(\ref{Eq:polarized_coherence}), which can be rewritten as
 \begin{equation}
 \label{Eq:inversion_IQUV}
\left[\begin{array}{l}
                      \!\!\! <E_x^nE_x^{m\star}>\!\!\!  \\
                      \!\!\! <E_x^nE_y^{m\star}> \!\!\! \\
                      \!\!\! <E_y^nE_x^{m\star}> \!\!\! \\
                      \!\!\! <E_y^nE_y^{m\star}> \!\!\!
                     \end{array} \right] 
                = 
\msgguy{I\stimes\frac{1}{2}}\left[\begin{array}{llll}
                      \!\!\!  1 & 1 & 0 & 0  \!\!\!  \\
                      \!\!\!  0 & 0 & 1 &  \msgref{i \!\!\!} \\
                      \!\!\!  0 & 0 & 1 & \msgref{\!\!\!\!-i \!\!\!} \\
                      \!\!\!  1 & \!\!\!\!-1 & 0 & 0  \!\!\! 
                     \end{array} \right] 
\left[\begin{array}{l}
                      \!\!\! V^{nm}_{\mathcal{I}} \!\!\!  \\
                      \!\!\! V^{nm}_{\mathcal{Q}} \!\!\! \\
                      \!\!\! V^{nm}_{\mathcal{U}} \!\!\! \\
                      \!\!\! V^{nm}_{\mathcal{V}} \!\!\!
                     \end{array} \right] 
.\end{equation}
The 4x4 matrix \msgguy{times $1/2$} is called $T_{I_2}$ and is used in Section~\ref{sec:propagation}. In what follows, the column on the left is called the coherence vector or the cross-coherence vector. When $n=m$, it is called the self-coherence vector. The determinant of the 4x4 matrix is equal to \msgref{$i/4$} and is nonzero. Note that the imaginary determinant encodes the $\pi/2$ phase difference between linear and circular analyzer bases and has no other physical meaning. Note also that for unpolarized sources $Q$, $U$, and $V$ are zero and $V^{nm}_{\mathcal{I}}$ is directly measured with the first and/or last correlation or the sum of the two, the problem then being scalar and not vectorial anymore. 
In the general case, the matrix can be inverted and the Stokes visibilities can be obtained from the components of the coherence matrix through the relation
 \begin{equation}
 \label{Eq:inversion_4x4_coherence_matrix}
\left[\begin{array}{l}
                      \!\!\! V^{nm}_{\mathcal{I}} \!\!\!  \\
                      \!\!\! V^{nm}_{\mathcal{Q}} \!\!\! \\
                      \!\!\! V^{nm}_{\mathcal{U}} \!\!\! \\
                      \!\!\! V^{nm}_{\mathcal{V}} \!\!\!
                     \end{array} \right] 
                = 
\frac{1}{I}\msgguy{\stimes}\left[\begin{array}{llll}
                      \!\!\!  1 & 0 & 0 & 1  \!\!\!  \\
                      \!\!\!  1 & 0 & 0 & \!\!\!\!-1  \!\!\! \\
                      \!\!\!  0 & 1 & 1 & 0  \!\!\!  \\
                      \!\!\!  0 & \msgref{\!\!\!\!-i}  & \msgref{i} & 0  \!\!\! 
                     \end{array} \right] 
\left[\begin{array}{l}
                      \!\!\! <E_x^nE_x^{m\star}>\!\!\!  \\
                      \!\!\! <E_x^nE_y^{m\star}> \!\!\! \\
                      \!\!\! <E_y^nE_x^{m\star}> \!\!\! \\
                      \!\!\! <E_y^nE_y^{m\star}> \!\!\!
                     \end{array} \right] 
.\end{equation}
The other particular case is when the source is only linearly polarized as circularly polarized light is often much fainter than linearly polarized light. In this case in which $V$ can be neglected with respect to $Q$ and $U$, $\mathcal{V}$ is set equal to zero and only three equations need to be inverted, which can be written as
 \begin{equation}
 \label{Eq:inversion_IQU}
\left[\begin{array}{l}
                      \!\!\!  \,\,\,\,\,\,\,\,\,\,\,\,\,\,\,\,\,\,\,<E_x^nE_x^{m\star}>\!\!\!  \\
                      \!\!\! <E_x^nE_y^{m\star}> + <E_y^n\msgguy{E_x^{m\star}}>\!\!\! \\
                      \!\!\! \,\,\,\,\,\,\,\,\,\,\,\,\,\,\,\,\,\,\,<E_y^n\msgguy{E_y^{m\star}}> \!\!\! \\
                     \end{array} \right] 
                = 
I\msgguy{\stimes}\frac{1}{2}\left[\begin{array}{llll}
                      \!\!\!  1 & 1 & 0  \!\!\!  \\
                      \!\!\!  0 & 0 & 2  \!\!\! \\
                      \!\!\!  1 & \!\!\!\!-1 & 0 \!\!\! 
                     \end{array} \right] 
\left[\begin{array}{l}
                      \!\!\! V^{nm}_{\mathcal{I}} \!\!\!  \\
                      \!\!\! V^{nm}_{\mathcal{Q}} \!\!\! \\
                      \!\!\! V^{nm}_{\mathcal{U}} \!\!\! 
                     \end{array} \right] 
,\end{equation}
the determinant of the 3x3 matrix now being $4$ and the middle coherence term being given by the difference of the Wollaston polarizer outputs after the action of a $22.5^o$ rotated half-wave plate. The visibilities are then given by
 \begin{equation}
 \label{Eq:inversion_3x3_coherence_matrix}
 \left[\begin{array}{l}
                      \!\!\! V^{nm}_{\mathcal{I}} \!\!\!  \\
                      \!\!\! V^{nm}_{\mathcal{Q}} \!\!\! \\
                      \!\!\! V^{nm}_{\mathcal{U}} \!\!\! 
                     \end{array} \right] 
               = 
\frac{1}{I}\msgguy{\stimes}\left[\begin{array}{llll}
                      \!\!\!  1 & 0 & 1  \!\!\!  \\
                      \!\!\!  1 & 0 & \!\!\!\!-1  \!\!\! \\
                      \!\!\!  0 & 1 & 0 \!\!\! 
                     \end{array} \right] 
\left[\begin{array}{l}
                      \!\!\!  \,\,\,\,\,\,\,\,\,\,\,\,\,\,\,\,\,\,\,<E_x^nE_x^{m\star}>\!\!\!  \\
                      \!\!\! <E_x^nE_y^{m\star}> + <E_y^n\msgguy{E_x^{m\star}}>\!\!\! \\
                      \!\!\! \,\,\,\,\,\,\,\,\,\,\,\,\,\,\,\,\,\,\,<E_y^n\msgguy{E_y^{m\star}}> \!\!\! \\
                     \end{array} \right] 
.\end{equation}
%
%

\section{Propagation of coherence in an optical interferometer}
\label{sec:propagation}
The propagation of waves in an optical system can be described with the formalism of 2x2 complex Jones matrices. If the input wave collected by a telescope is denoted as $\vect{E}$ and the optical system to transport the collected beam down to the beam combination point is described by the matrix $J$, then the exit beam reads as
\begin{equation}
\vect{\tilde{\!\!E}} = J.\vect{\!\!E}
.\end{equation}
The $\tilde{}$ symbol is used for propagated quantities in this paper. Using Eq.(\ref{Eq:coherence_product}) for the definition of the coherence matrix from the wave vectors, one gets that the coherence matrix, $C^{nm}$, is transported as the following propagated coherence matrix across the optical system:
\begin{equation}
\tilde{C}^{nm}=J^n C^{nm}J^{m\dag}
.\end{equation}
The above equation is known as the radio interferometer measurement equation (RIME), in for example \citet{Smirnov2011}, and was first introduced by \citet{Hamaker1996}. The equivalent of Eq.(\ref{Eq:polarized_coherence}) after propagation of the beams is written as
 \begin{equation}
 \label{Eq:propagated_coherence}
                \tilde{C}^{nm}= \frac{I}{2}V^{nm}_{\mathcal{I}}J^n.J^{m\dag}+\frac{I}{2}J^n.\left[\begin{array}{ll}
                      \!\!\!\,\,\,\,\,\,\,\,\,\,V_{\mathcal{Q}}^{nm}& \msgref{V_{\mathcal{U}}^{nm}+iV_{\mathcal{V}}^{nm}} \!\!\!  \\
                      \!\!\!\msgref{V_{\mathcal{U}}^{nm}-iV_{\mathcal{V}}^{nm}}&\,\,\,\,\,\,\,-V_{\mathcal{Q}}^{nm}\!\!\!
                     \end{array} \right].J^{m\dag}
.\end{equation}
Since in the radio domain the effect of instrumentation on the polarization can be more easily described by a few complex factors, the Jones propagation matrices are quickly expanded as the product of a few simple matrices at this stage.

As in Eq.(\ref{Eq:polarized_coherence}), the equation is written splitting the classical part with the source visibility, $V^{nm}_{\mathcal{I}}$, from the part with the Stokes visibilities, which can be understood as a bias if they are neglected. \msgref{This equation therefore has a major consequence: even in the case of an unpolarized source, the propagated coherence matrix may have nonzero antidiagonal terms, which should only exist for polarized sources. It therefore exhibits some polarization crosstalk, which creates biases and ghost polarized visibilities that need to be subtracted in the calibration process.} 

This relation can be rearranged in \msgguy{a system of equations}:
 \begin{equation}
\left[\begin{array}{l}
                      \!\!\! <\tilde{E}_x^n\tilde{E}_x^{m\star}>\!\!\!  \\
                      \!\!\! <\tilde{E}_x^n\tilde{E}_y^{m\star}> \!\!\! \\
                      \!\!\! <\tilde{E}_y^n\tilde{E}_x^{m\star}> \!\!\! \\
                      \!\!\! <\tilde{E}_y^n\tilde{E}_y^{m\star}> \!\!\!
                     \end{array} \right] 
                = 
\msgguy{I {\stimes}}T^{nm}
\left[\begin{array}{l}
                      \!\!\! V^{nm}_{\mathcal{I}} \!\!\!  \\
                      \!\!\! V^{nm}_{\mathcal{Q}} \!\!\! \\
                      \!\!\! V^{nm}_{\mathcal{U}} \!\!\! \\
                      \!\!\! V^{nm}_{\mathcal{V}} \!\!\!
                     \end{array} \right] 
.\end{equation}
In what follows, the 4x1 coherence and visibility column vectors of the above equation are denoted as $\tilde{\mathcal{C}}^{nm}$ and $\mathcal{V}^{nm}$ and the following equation is written:
 \begin{equation}
 \label{Eq:inversion_IQUV_prop}
\tilde{\mathcal{C}}^{nm} = \msgguy{I\stimes}T^{nm}. \mathcal{V}^{nm}
.\end{equation}
A similar expression can be derived with a 3x3 matrix in the case in which circular polarization can be neglected. The transfer matrix, $T^{nm}$, expression  is the following (splitting the 4x4 matrix into two lines):
 \begin{equation}
 \begin{split}
		T^{nm}=\msgguy{\frac{1}{2}}\left[\begin{array}{ll}
			\!\!\![J^{n}_{11}J^{m\star}_{11}+J^{n}_{12}J^{m\star}_{12}] & \!\![J^{n}_{11}J^{m\star}_{11}-J^{n}_{12}J^{m\star}_{12}]  \\
			\!\!\![J^{n}_{11}J^{m\star}_{21}+J^{n}_{12}J^{m\star}_{22}] & \!\![J^{n}_{11}J^{m\star}_{21}-J^{n}_{12}J^{m\star}_{22}]  \\
			\!\!\![J^{n}_{21}J^{m\star}_{11}+J^{n}_{22}J^{m\star}_{12}] & \!\![J^{n}_{21}J^{m\star}_{11}-J^{n}_{22}J^{m\star}_{12}]    \\
			\!\!\![J^{n}_{21}J^{m\star}_{21}+J^{n}_{22}J^{m\star}_{22}] & \!\![J^{n}_{21}J^{m\star}_{21}-J^{n}_{22}J^{m\star}_{22}]    
\end{array} \right. \\
&  \!\!\!\!\!\!\!\!\!\!\!\!\!\!\!\!\!\!\!\!\!\!\!\!\!\!\!\!\!\!\!\!\!\!\!\!\!\!\!\!\!\!\!\!\!\!\!\!\!\!\!\!\!\!\!\!\!\!\!\!\!\!\!\!\!\!\!\!\!\!\!\!\!\!\!\!\!\!\!\!\!\!\!\!\!\!\!\!\!\!\!\!\!\!\!   \!\!\!\!
\left.\begin{array}{ll}
\,\,\,\,\,\,\,\,[J^{n}_{11}J^{m\star}_{12}+J^{n}_{12}J^{m\star}_{11}] & \msgref{ \!\!i[J^{n}_{11}J^{m\star}_{12}-J^{n}_{12}J^{m\star}_{11}] \!\!\!} \\
\,\,\,\,\,\,\,\,[J^{n}_{11}J^{m\star}_{22}+J^{n}_{12}J^{m\star}_{21}] & \msgref{\!\!i[J^{n}_{11}J^{m\star}_{22}-J^{n}_{12}J^{m\star}_{21}] \!\!\! }\\
\,\,\,\,\,\,\,\,[J^{n}_{21}J^{m\star}_{12}+J^{n}_{22}J^{m\star}_{11}] & \msgref{\!\!i[J^{n}_{21}J^{m\star}_{12}-J^{n}_{22}J^{m\star}_{11}] \!\!\! }\\
\,\,\,\,\,\,\,\,[J^{n}_{21}J^{m\star}_{22}+J^{n}_{22}J^{m\star}_{21}] & \msgref{\!\!i[J^{n}_{21}J^{m\star}_{22}-J^{n}_{22}J^{m\star}_{21}] \!\!\! }
\end{array} \right]
\end{split}
\end{equation}
\msgguy{The columns of $T^{nm}$ can be computed very conveniently with the Pauli matrices, $\sigma_0, \sigma_1, \sigma_2, \sigma_3$  (with $\sigma_0 = I_2$). As matter of fact, the first column is obtained by reorganizing the terms of $\mathrm{Col}_1 = J^n \sigma_0 J^{m\dagger}$ into a single column with on the first line Col$_1$[1,1], on the second line Col$_1$[1,2], on the third line Col$_1$[2,1], and on the last line Col$_1$[2,2]. The same is then done for $\mathrm{Col}_2 = J^n \sigma_3 J^{m\dagger}$, $\mathrm{Col}_3 = J^n \sigma_1 J^{m\dagger}$, and $\mathrm{Col}_4 = J^n \sigma_2^{\star} J^{m\dagger}$.} \\
The $T_{I_2}$ matrix introduced in Eq.(\ref{Eq:inversion_IQUV}) of Section~\ref{sec:formalism} is equal to the transfer matrix, $T^{nm}$, in the special case in which $J_n=J_m=I_2$. And one gets the 3x3 matrix of Eq.(\ref{Eq:inversion_IQU}) by suppressing the fourth column and adding the second and third lines. For $n=m$, $\mathcal{S}=I\mathcal{V}^{nn}$ is the Stokes parameter vector (it is independent of $n$) and $T_{I_2}^{-1}T^{nn}$ is the Mueller matrix describing the propagation of the Stokes parameters for beam $n$. \msgguy{By analogy, I defined the generalized Mueller matrix, $M^{nm}=T_{I_2}^{-1}T^{nm}$, to describe the propagation of the Stokes visibilities in the optical system according to the equation}
\begin{equation}
\label{Eq:generalized_Mueller}
\msgguy{\tilde{\mathcal{V}}^{nm} = M^{nm}.\mathcal{V}^{nm} }
.\end{equation}
%

%
%
\section{Calibration of visibilities of polarized beams}
\label{sec:calibration}
\subsection{Visibility moduli}
\label{sec:visibility_modulus}
\subsubsection{Case of polarized sources}
\label{sec:polarized_sources}
In this section, it is assumed that the interferometer is single-mode, i.e., that the beams are filtered with single-mode fibers before beam combination ensuring that phase fluctuations in the pupil plane are traded against intensity fluctuations which can be calibrated, as demonstrated by the FLUOR instrument \citep{Foresto1997,Perrin1998}.
In practice, the Jones matrices, $J^n$, are measured to within a complex factor, which can be called a gain and denoted as $g_n$. \msgguy{Eq.(\ref{Eq:generalized_Mueller})} is then modified as
\begin{equation}
 \label{Eq:inversion_IQUV_prop_gain_Mueller}
\msgguy{\tilde{\mathcal{V}}^{nm} = g_n g_m^\star M^{nm}. \mathcal{V}^{nm} }
,\end{equation}
\msgguy{or, equivalently, since $T^{nm}=T_{I_2}.M^{nm}$,} Eq.(\ref{Eq:inversion_IQUV_prop}) \msgguy{can be} modified as
\begin{equation}
 \label{Eq:inversion_IQUV_prop_gain}
\tilde{\mathcal{C}}^{nm} = \msgguy{I}g_n g_m^\star T^{nm}. \mathcal{V}^{nm}
.\end{equation}
The same applies for the self-coherence measured for each beam, which here in the case of beam $n$ is
\begin{equation}
\label{Eq:phot_gain}
\tilde{\mathcal{P}}^{n} = |g_n|^2 T^{nn} .\mathcal{S}
.\end{equation}
$\mathcal{S}\!=\!I\mathcal{V}^{nn}\!\!=\!I\mathcal{V}^{mm}$ is the Stokes parameter vector. Some extra gains may need to be introduced depending on the type of beam combination and the calibration protocol, but these are calibrated on an internal source and are not mentioned here to keep formulas as simple as possible. The self-coherence for beam $n$ is denoted as $\mathcal{P}^{n}$, signifying {photometry for beam $n$} for a reason that will become obvious in the next section. The equation is modified by multiplying the two members by the inverse of matrix $T^{nn}$ and by taking the scalar product with $\mathcal{S}$:
\begin{equation}
\label{Eq:phot_gain}
\mathcal{S}^{\dag}.\left.T^{nn}\right.^{-1}.\tilde{\mathcal{P}}^{n} =|g_n|^2  \mathcal{S}^{\dag}.\mathcal{S}
.\end{equation}
$\mathcal{S}^{\dag}.\mathcal{S}$ takes the simple expression
\begin{equation}
\mathcal{S}^{\dag}.\mathcal{S} = I^2+Q^2+U^2+V^2= (1+P^2)I^2
;\end{equation}
 hence, the value of the square of the modulus of the gain,
\begin{equation}
 \label{Eq:phot_gain}
 |g_n|^2 =  \frac{\msgguy{1}}{(1+P^2)I^2}\mathcal{S}^{\dag}.\left.T^{nn}\right.^{-1}.\tilde{\mathcal{P}}^{n}
.\end{equation}
The same is obtained for beam $m$. \msgguy{Eq.(\ref{Eq:inversion_IQUV_prop_gain_Mueller})} can be inverted to
\begin{equation}
\msgguy{\mathcal{V}^{nm} = \frac{1}{g_n g_m^\star}  \left.M^{nm}\right.^{-1} .\tilde{\mathcal{V}}^{nm} }
.\end{equation}
And one gets for the moduli of the Stokes visibilities (the modulus applied to the vector is the modulus of its components):
\begin{equation}
\label{Eq:calibrated_Stokes_visibility}
\msgguy{|\mathcal{V}^{nm}| = \frac{I^2(1+P^2)| \left.M^{nm}\right.^{-1}. \tilde{\mathcal{V}^{nm}}|}{{\sqrt{\mathcal{S}^{\dag}.T^{nn\,-1}.\tilde{\mathcal{P}}^{n}}\sqrt{\mathcal{S}^{\dag}.T^{mm\,-1}.\tilde{\mathcal{P}}^{m}}}} }
.\end{equation}
Note that the measurement of the Stokes parameters usually leads to $\mathcal{S}$ normalized by the measured total intensity. \msgguy{In the same vein, either one measures the coherent Stokes flux, $I\tilde{\mathcal{V}^{nm}}$, and the self-coherent flux, $\tilde{\mathcal{P}}^{m}$, or the Stokes visibilities, $\tilde{\mathcal{V}^{nm}}$, and the normalized self-coherent flux, $\tilde{\mathcal{P}}^{m}/I$. In either case,} the modulus of the Stokes visibility \msgguy{is} independent of the absolute value of the total intensity, $I$, in the above formula. 
\subsubsection{Case of unpolarized sources}
\msgguy{In absence of polarization crosstalk, i.e., when $<\!\!\!\tilde{E}_x^n\tilde{E}_y^{m\star}\!\!\!>=<\!\!\!\tilde{E}_y^n\tilde{E}_x^{m\star}\!\!\!> =0$, t}he above equations take a simpler form in the case in which the degree of polarization is $P=0$ or, equivalently, when the Stokes parameters $Q=U=V=0$. \msgguy{This leads} to scalar values for the visibilities\msgguy{,} coherence and self-coherence as in this case the visibility is only $V^{nm}_\mathcal{I}$ and the first and last element of the self-coherence vector are the only nonzero terms and can be summed to obtain the total intensity detected by the instrument for a given beam, and hence the term of photometry in this case. The same applies as well for the cross-coherence and one sums the first and last terms of the coherence vector to get the total coherent intensity $\tilde{I}^{nm}=\tilde{I}^{nm}_x+\tilde{I}^{nm}_y$. As a particular case of the above equations, one gets for the coherent intensity
\begin{equation}
\tilde{I}^{nm} = \frac{I}{2}V^{nm}_\mathcal{I}\left(\sum_{i,j}J^{n}_{ij}J^{m\star}_{ij}\right) \stimes g_n g_m^\star
,\end{equation}
and for the photometries
\begin{equation}
\tilde{P}^{n,m} = \frac{I}{2}\left(\sum_{i,j}|J^{n,m}_{ij}|^2\right) \stimes |g_{n,m}|^2
.\end{equation}
So that the equivalent of Eq.(\ref{Eq:calibrated_Stokes_visibility}) is written as
\begin{equation}
|V^{nm}_\mathcal{I}| = \frac{|\tilde{I}^{nm}|}{\sqrt{\tilde{P}^{n} \tilde{P}^{m}}}\stimes\frac{\sqrt{\sum_{i,j}|J^{n}_{ij}|^2\sum_{i,j}|J^{m}_{ij}|^2}}{|\sum_{i,j}J^{n}_{ij}J^{m\star}_{ij}|}
.\end{equation}
The ratio after the \msgguy{times} sign is the reciprocal of  $\sqrt{\kappa_n \kappa_m}$ of \citet{Foresto1997}.\\

\msgref{In case of polarization crosstalk, as in \citet{Buscher2009}, these expressions need to be debiased using the 4D Stokes visibility formalism presented in Section~\ref{sec:polarized_sources} as ghost polarized visibilities are produced by the optical system.}

\subsection{Visibility phases \msgref{and closure phases}}
\label{sec:visibility_phases}
\msgguy{Eq.(\ref{Eq:inversion_IQUV_prop_gain_Mueller})} can be modified multiplying both sides by \msgguy{$M^{nm\,-1}$}, yielding
\begin{equation}
\msgguy{M^{nm\,-1}.\tilde{\mathcal{V}}^{nm} = g_n g_m^\star  \mathcal{V}^{nm} }
.\end{equation}
\msgref{This equation shows how to get access to visibilities from which the effects of polarization crosstalks have been removed. The remaining phase bias due to the gain product, $g_n g_m^\star $, can be calibrated out by subtracting the phase observed on a reference source in the case of a dual-field instrument such as GRAVITY.} \\

\msgref{For single-field instruments, the above relation allows one to obtain an unbiased closure phase quantity.}  Taking the Hadamard product of the above column vector for baselines $nm$, $ml$, and $ln$, one gets
\begin{equation}
\begin{split}
\msgguy{(M^{nm\,-1}\!\!.\tilde{\mathcal{V}}^{nm})\odot (M^{ml\,-1}\!\!.\tilde{\mathcal{V}}^{ml})\odot (M^{ln\,-1}\!\!.\tilde{\mathcal{V}}^{ln})\!=} & \\
& \!\!\!\!\! \!\!\!\!\! \!\!\!\!\! \!\!\!\!\! \!\!\!\!\! \!\! \! \msgguy{|g_n g_m g_l|^2  \mathcal{V}^{nm} \odot \mathcal{V}^{ml} \odot \mathcal{V}^{ln} }
\end{split}
,\end{equation}
and therefore the closure phase relation for all Stokes visibilities:
\begin{equation}
\begin{split}
\msgguy{\mathrm{Arg}\!\left((M^{nm\,-1}\!\!.\tilde{\mathcal{V}}^{nm})\!\odot \!(M^{ml\,-1}\!\!.\tilde{\mathcal{V}}^{ml})\!\odot \!(M^{ln\,-1}\!\!.\tilde{\mathcal{V}}^{ln})\right)\!= }& \\
& \!\!\! \!\!\!\!\! \!\!\!\!\! \!\!\! \! \!\!\! \!\!\! \! \mathrm{Arg} \!\left(\mathcal{V}^{nm}\!\! \odot \!\mathcal{V}^{ml} \!\!\odot \! \mathcal{V}^{ln}\right)
\end{split}
.\end{equation}
%
%

\section{Examples of polarized waves and of interferometer polarizing properties}
\label{sec:examples}
In this section some specific cases are given to illustrate the above theory and also make a link with some previous publications.

\subsection{Effect of beam rotation and birefringence}
The process of detection is a major difference between radio interferometry and optical interferometry as in the latter case only intensity can be detected (except for wavelengths beyond 10$\,\mu$m where heterodyne techniques can be employed, as was the case with the Infrared Spatial Interferometer \cite{Hale2000}) and therefore beams must be transported down to the beam combination point. A consequence is the complexity of beam trains, which must include, on top of that, delay lines to compensate for the external variable optical path difference. Modern optical interferometers are almost perfectly symmetric but still have some imperfections, causing small contrast losses. In \citet{Perrin2024b}, I showed that, at least in the case of the VLTI, the Jones matrices of beam trains can be approximated with a good accuracy by what I called quasi-unitary matrices of the form
\begin{equation}
\label{eq:quasi_unitary}
U^q(\theta,\psi,\varphi_1,\varphi_2,\tau_1,\tau_2) = R_\theta \, R_\psi \, \Delta(\tau_1 e^{i\varphi_1},\tau_2 e^{i\varphi_2}) \, R_{-\psi}
,\end{equation}
where $R_\theta$ is a global rotation of angle $\theta$, and $\Delta$ is a diagonal matrix describing the polarizing properties of the system (diattenuation $\tau_2 - \tau_1$ and birefringence $\varphi_2 - \varphi_1$) described for its neutral axis rotated by the angle $\psi$ with respect to the reference axes. This case was studied by \citet{Perraut1997} for the GI2T interferometer considering identical neutral axes orientations in all  beams but with different phases and coefficients and different global rotations using the following quasi-unitary matrix for beam $i$:  $J_i=U^q(\theta_i,0,\Phi_{s,i},\Phi_{p,i},R_{s,i},R_{p,i})$. In a simpler version, they considered the effects of differential beam rotation only ($J_i=U^q(\theta_i,0,0,0,1,1)$) and of differential birefringence only ($J_i=U^q(0,0,\phi_{s,i},\phi_{p,i},1,1)$), first assuming no attenuation and then mixing all effects \citep{Perraut1996}. \msgref{\citet{Shuai2025} have modeled the polarization properties of CHARA by decomposing the beam train in groups of optical elements, each described by unitary or quasi-unitary Jones matrices with beam rotation, diattenuation, and retardance, and applied their model to observations of polarized sources.} In \citet{Perrin2025}, I also considered quasi-unitary matrixes to discuss biases produced by retardance and diattenuation in the measurements of long-baseline interferometers for point-like sources.

\subsection{Case of symmetric interferometers without diattenuation}
For such interferometers, all the Jones matrices are \msgguy{exactly proportional to unitary matrices and are} of the type: $J_i=U^q(\theta,0,\varphi_{1},\varphi_{2},|g|,|g|) = |g| U(\theta,0,\varphi_{1},\varphi_{2})$, independently of $i$ and where $U(\theta,0,\varphi_{1},\varphi_{2})$ is a unitary matrix (all unitary matrices can be written as the product of rotation matrices with a diagonal unitary matrix\msgguy{, see \citet{Perrin2024b}}). Eq.(\ref{Eq:propagated_coherence}) becomes in this case
 \begin{equation}
                \tilde{C}^{nm}= |g|^2\frac{I}{2}V^{nm}_{\mathcal{I}}I_2+|g|^2\frac{I}{2}U.\left[\begin{array}{ll}
                      \!\!\!\,\,\,\,\,\,\,\,\,\,V_{\mathcal{Q}}^{nm}& \msgref{V_{\mathcal{U}}^{nm}+iV_{\mathcal{V}}^{nm} \!\!\!}  \\
                  \msgref{    \!\!\!V_{\mathcal{U}}^{nm}-iV_{\mathcal{V}}^{nm} }&\,\,\,\,\,\,\,-V_{\mathcal{Q}}^{nm}\!\!\!
                     \end{array} \right].U^{\dag}
.\end{equation}
The effect on $V^{nm}_{\mathcal{I}}$ of the polarization properties of the interferometer cancels but it does not cancel for the polarized part of the visibilities. This shows that even when an interferometer is perfectly symmetric and without diattenuation, the optical beam properties need to be taken into account to calibrate the visibilities, unless the degree of polarization of the source is null.

\subsection{Case of a polarized point source}
Here, the polarized emitter is a point source, in addition to an unpolarized spatial intensity distribution. The Stokes visibilities are constant and written as $Q/I$, $U/I$ and $V/I$. Hence the expression of Eq.(\ref{Eq:propagated_coherence}) in this case is
 \begin{equation}
                \tilde{C}^{nm}= \frac{I}{2}V^{nm}_{\mathcal{I}}J^n.J^{m\dag}+\frac{1}{2}J^n.\left[\begin{array}{ll}
                      \!\!\!\,\,\,\,\,\,Q& \msgref{\!\!\!U+iV \!\!\!}  \\
                     \msgref{ \!\!\!U-iV } &-Q\!\!\!
                     \end{array} \right].J^{m\dag}
.\end{equation}
As in traditional interferometry, the visibility of the unpolarized source is offset by the visibility of the polarized point source but also with an additional term that is proportional to the Stokes parameters of the point source, the second term in the above equation.

\subsection{Case of sources with uniform polarization properties}
It is now supposed that the polarized emitter is an extended source but with uniform polarization properties across the object. In this case, the Stokes visibilities (except for $I$) are proportional to the Fourier transform of the support of the object, $A(\alpha,\delta)$, whose integral is the area covered by the object, $\mathcal{A}$, and are written as
\begin{equation}
 \left\{\begin{array}{@{}l@{}}
V_{\mathcal{Q}}(u,v) \,= Q\mathrm{FT}[A](u,v)/(\mathcal{A}I)\\ 
V_{\mathcal{U}}(u,v) = U\mathrm{FT}[A](u,v)/(\mathcal{A}I)\\ 
V_{\mathcal{V}}(u,v) = V\mathrm{FT}[A](u,v)/(\mathcal{A}I)\\ 
\end{array}\right.
.\end{equation}
Eq.(\ref{Eq:propagated_coherence}) then becomes
 \begin{equation}
                \tilde{C}^{nm}= \frac{I}{2}V^{nm}_{\mathcal{I}}J^n.J^{m\dag}+\frac{1}{2}\frac{\mathrm{FT}[A]}{\mathcal{A}}J^n.\left[\begin{array}{ll}
                      \!\!\!\,\,\,\,\,\,Q& \msgref{\!\!\!U+iV \!\!\! } \\
                     \msgref{ \!\!\!U-iV } &-Q\!\!\!
                     \end{array} \right].J^{m\dag}
.\end{equation}
The more extended the object, the narrower the visibility function, $\frac{\mathrm{FT}[A]}{\mathcal{A}}$. This function becomes negligible past a spatial frequency that is proportional to the reciprocal of size of the object. Beyond this value, the cross-coherence is almost not biased by the Stokes visibilities but the visibilities are biased by the source polarization after the photometric calibration if this term is neglected.

\section{Conclusions}
\label{sec:conclusion}
Based on previous results on aperture synthesis in radio astronomy, a full and consistent formalism for long baseline interferometry in the optical domain, taking into account the polarization properties of the interferometer and the polarimetric characteristics of the source, is established. Spatial coherence is shown to be linked to the Stokes visibilities through a simple matrix relationship for each baseline of the interferometer. The Stokes visibilities are a generalization of the classical visibility function but they are applied to the Stokes parameters, and therefore they describe the spatial distribution of the polarimetric characteristics of the astronomical source. \msgguy{A generalized Mueller matrix is introduced to describe the propagation of the Stokes visibilities across the instrument as an analogy with the Mueller matrix that describes the propagation of the Stokes parameters in the single-beam case.} The matrix describes the polarization properties of the interferometer for a given baseline using the Jones formalism. This simple formula shows that both the polarimetry characteristics of the source and those of the instrument need to be taken into account to calibrate the visibilities. \msgref{It is in particular shown that complex visibilities and derived quantities such as the squared moduli, differential phases, and closure phases need to be debiased from polarization crosstalk, even when the source is not polarized as in this case ghost polarized visibilities are produced by the instrument and need to be removed in the calibration process using the generalized Mueller matrix.} The \msgguy{matrix formalism therefore} provides a \msgguy{method} of deriving \msgguy{the object} Stokes visibilities from the observed  \msgguy{visibilities}, and therefore yields constraints on the spatial distribution of the Stokes parameters. The principles of the calibration in the case of a single-mode interferometer are \msgguy{presented} in the paper. 

\begin{acknowledgements}
The author is thankful to Steve Ridgway, Karine Perraut and Oleg Smirnov for their careful reading and their suggestions to improve the article. \msgguy{The author also thanks the anonymous referee for the very useful comments that led to the last version of the paper}.
\end{acknowledgements}

\bibliographystyle{aa} 
\bibliography{references_pol_interferometry}

\end{document}